\documentstyle[preprint,aps]{revtex}
\begin{document}
\draft
\newcommand{\be}{\begin{equation}}
\newcommand{\ee}{\end{equation}}
\newcommand{\bea}{\begin{eqnarray}}
\newcommand{\eea}{\end{eqnarray}}
\title{Periodic Orbits in a Simple Ray-Splitting System}
\author{Debabrata Biswas \thanks{email:biswas@kaos.nbi.dk}
\thanks{on leave from Theoretical Physics Division, Bhabha Atomic Research
Centre, Bombay 400 085}}
\address{Center for Chaos and Turbulence Studies,
Niels Bohr Institute \\
Blegdamsvej 17, 
Copenhagen $\O$, Denmark}
\maketitle
\vskip .1 in
\begin{abstract}

\par We study  ray dynamics in a square billiard that allows mode
conversion and is parametrised by $\kappa$, the ratio of the velocities
of the two modes. At $\kappa \rightarrow 1^+$, conversion occurs
at every reflection and periodic orbits proliferate exponentially.
As $\kappa$ increases beyond $\sqrt{2}$, the collection of daughter
rays explore only three momentum directions and mode conversion is
progressively inhibited. We provide an algorithm for determining
periodic orbits when $\kappa > \sqrt{2}$ and show numerically
that exponential proliferation persists around $\kappa \simeq \sqrt{2}$
but as $\kappa$ increases, a crossover to sub-exponential behaviour
occurs for short periods. We discuss these results in the light
of conservation laws.

\end{abstract}

\vskip 0.45 in
\pacs{PACS numbers: 05.45.+b, 03.65.Sq}

\nopagebreak

\section{Introduction}
\label{sec:intro}

\par Billiards are useful examples of dynamical systems and display
the wide variety of phenomena associated with Hamiltonian flows.
The ray equations commonly considered follow from a short wavelength
expansion of the Schr\"{o}dinger equation with Dirichlet (or Neumann)
boundary conditions. A particle thus moves freely between collisions
at the boundary where it suffers specular reflection
and one can observe regular or chaotic motion
depending on the shape of the boundary. As an example, a square 
billiard generates regular dynamics that is restricted to a 
torus in phase space due to the existence of two well behaved 
constants of motion. In contrast, generic trajectories in the 
stadium shaped billiard explore the entire constant energy
surface and hence the system is ergodic. Moreover, orbits
that are nearby initially move exponentially apart with time
and hence the system is said to be metrically chaotic. 

\par These differences also show up in the proliferation
rates of periodic solutions. In case of regular motion, periodic
orbits exist in 1-parameter families and their number increases
quadratically with time period. In contrast, chaotic dynamics 
is accompanied by an exponential proliferation of periodic
orbits, a phenomenon referred to as topological chaos.
 
\par The manner in which periodic orbits organise themselves
in closed systems is strongly linked to the existence of
sum rules arising from conservation laws. For example, the
fact that a particle never escapes implies that for 
chaotic systems \cite{PCBE,HO}:

\be \langle \sum_p \sum_{r=1}^{\infty} {T_p \delta(t-rT_p)\over 
\left |{ \det({\bf 1} - {{\bf J}_p}^r)} \right |} \rangle = 1\label{eq:hbolic}
\ee

\noindent
where the summation over p refers to all primitive periodic
orbits, $T_p$ is time period, ${\bf J}_p$ is the stability
matrix evaluated on the orbit and the symbol $\langle . \rangle$
denotes the average value of the expression on the
left. Since the periodic orbits are
unstable and isolated, $\left | \det({\bf 1} - {{\bf J}_p}^r) \right | 
\simeq e^{\lambda_p rT_p}$, where $\lambda_p$ is the Lyapunov
exponent of the orbit. The exponential proliferation of orbits
is thus implicit in eq.~(\ref{eq:hbolic}).  

\par For the square billiard though, the appropriate sum rule is \cite{HO,db1}:

\be \langle {\sum_{M} \sum_{N}} {1\over (\pi/2)}{4A\delta(t - T_{M,N})
\over v^2T_{M,N}} \rangle = 1 \label{eq:consv} \ee

\noindent
where $v$ is the velocity, $T_{M,N}$ is the time period of a periodic orbit
with winding numbers (M,N) and $A$ is the area of the billiard.
The quadratic law  for the number of periodic orbits having time
period less than $T$ is thus contained in eq.~(\ref{eq:consv}).

\par In both cases, the stability of periodic orbits leads
to the respective proliferation laws. However, there can exist
situations where rays split up at the boundaries of billiards
and thus suffer a decay in intensity. The basic conservation 
law then demands that periodic orbits should proliferate faster
than usual to compensate this loss.

\par We explore here the case of a square billiard
that admits ray-splitting and is parametrised by $\kappa$, the
ratio of velocities of the two modes. For $\kappa \rightarrow 1^+$,
mode conversion 
occurs at every reflection from the boundary and 
daughter rays multiply as $2^k$ where $k$ denotes the number
of reflections. Moreover, the collection of daughter rays 
(from a single parent) can access an increasing number of momenta 
directions with time. As $\kappa$ increases, the range
of angle in which conversion can occur at adjacent edges
decreases and consequently the average number of daughter 
rays produced with every reflection decreases. 
For $\kappa \geq \sqrt{2}$, the possibility of mode
conversion is restricted further and daughter rays can access only 
three momentum directions. Simultaneously, the range 
of angles at which mode conversion can occur shrinks
as $\kappa$ increases beyond $\kappa_c = \sqrt{2}$.
In the limit $\kappa \rightarrow \infty$ conversions
are not allowed and the system is then a normal billiard 
without ray-splitting.

\par In the setting described above, we study the proliferation
law of periodic paths. In particular, we explore the 
region $\kappa \geq \sqrt{2}$ and provide an algorithm
for determining periodic paths. We find that for 
$\kappa = 1.429\; (> \kappa_c)$, the proliferation law remains exponential
but as $\kappa$ increases further, mode conversion is
inhibited and the proliferation law shows a crossover
to sub-exponential behaviour for short periods. 
 
\par Before discussing these in detail, it is important to
note that periodic orbits form the skeleton on which
modern semiclassics is built \cite{MCG} and are used to 
understand the spectrum in quantum systems. In
elastodynamics where mode conversion does occur,
studies of the statistical properties of the 
resonance spectrum are influenced to a large extent
by these considerations \cite{sornette,mark}. 
In fact, a fourier transform 
of the resonance spectrum provides evidence of the role
of periodic orbits in elastodynamics \cite{sornette}. The present 
study is thus significant in that it provides the 
first systematic computation of periodic orbits
and underscores its complex organisation even in
simple geometries.

\par In the following section, we describe in more detail the
system that we study and analyse the dynamics of the 
converted rays. Section \ref{sec:periodic} deals with the
algorithm for determining periodic orbits and our numerical
results are discussed in section \ref{sec:numerics}.

\section{A Simple Ray-Splitting System}
\label{sec:system}

\par Ray-splitting is a common phenomenon in geometrical
optics and can be observed in several other situations
\cite{ott,bertie,bertie1}. The propagation of elasto-mechanical
waves in solids is an example
that is of much current interest \cite{sornette,mark} and  
we study here the short wave-length limit of this problem
for the isotropic case 
in two dimensions. For small
displacements, the wave motion is governed by the
Navier equation~\cite{book} : 

\be \label{eq:navier}
\mu \nabla^2{\bf u} + (\lambda + \mu)\nabla \nabla \cdot {\bf u}
= \rho \partial^2 {\bf u}/{\partial t^2} 
\ee

\noindent
where {\bf u} is the displacement, $\lambda $ and $\mu $ are 
the two Lam\'{e} constants \cite{fnote1} and $\rho$ is the
density. It is common to 
express the displacement as a sum of two parts generated
respectively by a scalar and a vector potential (${\bf u} = 
{\bf u}_1 + {\bf u}_2 $ where ${\bf u}_1 = \nabla \phi$ and
${\bf u}_2 = \nabla \times {\bf B}$) so that eq.~(\ref{eq:navier}) separates
into two second-order equations \cite{book}: 

\begin{eqnarray}
\nabla^2\phi    & = & {1\over c_P^2} \partial^2 \phi/{\partial t^2} \label{eq:graf1} \\
\nabla^2{\bf B} & = & {1\over c_S^2} \partial^2 {\bf B}/{\partial t^2} \label{eq:graf2}
\end{eqnarray} 

\noindent 
where $c_P^2 = (\lambda + 2\mu )/\rho $ and $c_S^2 = \mu/\rho$.
The medium thus has two natural velocities, $c_P$ and $c_S$
where the subscripts $P$ and $S$ refer to the pressure (longitudinal)
and shear (transverse) waves respectively. The 
two waves  interact only at the boundaries where they may 
suffer mode conversion.

\par Consider for example an $S$ or a $P$ wave incident on 
a planar stress-free (tractionless) boundary.
The reflected part consists in general of both an $S$  
and a $P$ wave and the reflection law involving an $S$ and a 
$P$ wave is given by Snell's relation \cite{book}: 
 
\be \cos(\theta_P) = \kappa \cos(\theta_S)  \label{eq:snell} \ee

\noindent
where the angles are measured with respect to the tangent at
the collision point on the boundary and $\kappa = c_P/c_S$. 
When the reflected and incident wave belong to the same 
type (both $S$ or both $P$), the angle of reflection equals 
the angle of incidence.
Note that eq.~\ref{eq:snell} implies the existence of a critical
angle, $\theta_C = \cos^{-1}(1/\kappa)$. 
For $\theta_S < \theta_C$, no conversion
can occur and the wave suffers only a specular reflection.  

\par The incident and reflected amplitudes are also
determined by the boundary conditions. For $S \rightarrow S + P$
process, and for the case where the $S$ wave polarisation lies
in the plane of incidence, the intensities carried away by the 
reflected $S$ and $P$ wave are \cite{book}:

\be I_{SS} = \left | {\sin (2\theta_S) \sin(2\theta_P) - \kappa^2
\cos^2(2\theta_S) \over \sin (2\theta_S) \sin(2\theta_P) + \kappa^2
\cos^2(2\theta_S)} \right |^2 \label{eq:inten}  \ee 

\noindent
and $I_{SP} = 1 - I_{SS}$ respectively. Similarly, for the 
$P \rightarrow S + P$ process, $I_{PP} = I_{SS}$ and 
$I_{SP} = I_{PS} $ provided again that the $S$ wave polarisation lies
in the plane of incidence. In general, the full 3-dimensional
problem requires a decomposition of the polarisation vector
into components normal and parallel to the plane of incidence.
The component in the
plane undergoes the process described above while the normal
component suffers no conversion.

\par The small-wavelength limit of this wave motion restricted to a 
plane can be treated as the relevant ray-tracing problem in 
2-dimensions. The geometry we consider here is a square
billiard of length, $L = \pi/2$. 
In a situation where no mode conversion occurs,
the square billiard is integrable and its periodic
orbits increase quadratically with length. The corresponding
Schr\"{o}dinger equation can be solved easily and the 
quantum spectrum is described exactly by periodic orbits.
The situation in elasdo-dynamics is however quite the
opposite and the only resonances known analytically
form a small fraction of the total number \cite{book}.
To the best of our knowledge, the ray-tracing problem
in this geometry has not been studied before and we 
proceed to understand that now.

\par The parameter range that is physically accessible
is $\kappa > 1$ and as an example we study the case 
$\kappa = 1.429$. 
Fig.~1 illustrates the conversions that can occur in this
system. For the sake of visualisation, it is convenient to 
consider an unfolded trajectory in the full plane generated by
reflections of the fundamental domain about its sides (see Fig. 2).
A reflected trajectory without conversion thus continues through the
boundary without any change in angle while a ray that has suffered
conversion is equivalent to a refracted ray that undergoes
a change in angle. A cell identical to the fundamental domain
can then be labelled by the winding numbers $(M,N)$  
where $2ML$ is the displacement of any point in the cell along 
the X-axis and $2NL$ along the Y-axis. Fig. 2 shows the cells
with winding numbers (1,1) and (2,2). 

\par Consider then a family of $S$ rays at an angle
$\theta_1$ with respect to the horizontal edge (X-axis, see Fig. 2).
For $\theta_1 < \pi/2 - \theta_C$, 
conversion to a $P$ ray can occur only at the 
vertical edge (Y-axis, see Fig. 2). The $P$-ray at an angle
$\theta_2$ (measured from the X-axis), can however
reconvert to an $S$ ray at both the vertical and the horizontal
edges. If it reconverts at the vertical edge, the resultant
reflected $S$-ray is again at an angle $\theta_1$ while 
if reconversion occurs at the horizontal edge,
the $S$-ray (which we denote by $S'$) is at an angle 
$\theta_3 > \theta_1$ (see Fig. 1). 
The first possibility requires no further analysis  
while for the $S$-ray at an angle $\theta_3$, reconversion
can occur only at the horizontal edge and hence
generates a $P$ ray at angle $\theta_2$. An identical
process occurs if $\theta_1 > \theta_C$ except
that edges are now interchanged. For $\pi/2 - \theta_C < \theta_1 <
\theta_c$ (see Fig. 1) and for $\theta_1 = 0$ or $\pi/2$, 
there is no conversion
and the orbit continues with the same intensity.
This exhausts all
possibilities so that there are only three 
directions $\{\theta_1,\theta_2,\theta_3\}$ that all rays
originating from a given parent ray can explore.

\par As $\kappa$ increases, the two branches in the map of
Fig.~1 move apart and the range of angles at which
conversion can occur at either the vertical or horizontal
edge shrinks. However, the scenario described above 
continues to hold whenever conversion does occur and
the daughter rays produced explore only three momenta
directions.  

\par As $\kappa$ decreases however, the branches in the map
(of Fig.~1) come closer and meet at 
$\kappa_c = \sqrt{2}$. As $\kappa$ decreases below $\kappa_c$,
there is a range of angles in which conversion  
occurs both at the horizontal and vertical edge as
shown in Fig.~(3). Consequently, the number of directions
accessible to all daughter rays from a single parent
increases as well. As $\kappa$ decreases further, the 
overlap increases and at $\kappa \rightarrow 1^+$, 
conversion can occur at both the horizontal and 
vertical edge for any initial angle. Thus rays split with
every reflection and the number of daughter rays grows
as $2^k$ where $k$ denotes the number of reflections.   
The collection of all daughter rays now explore 
an increasing number of momentum directions in sharp
contrast to the case when $\kappa > \kappa_c$.

\par Note that in all cases, the orbits are marginally
unstable as in the case when no conversion occurs. This
can be verified by linearising the neighbourhood and 
looking at the eigenvalues of the Jacobian matrix. 
Periodic rays thus occur in families and their extent
is limited by the vertices where adjacent parallel
rays meet a horizontal and a vertical edge respectively
and convert differently.

\section{Periodic Orbits}
\label{sec:periodic}

\par We now turn to a study of periodic orbits in such
a system and provide an algorithm for determining them
when $\kappa > \kappa_c$.  

\par We first note that unlike a normal square  
billiard, each set of winding numbers ($M,N$) can correspond
to more than one periodic solution. The
total length, $l_S$, of all $S$ segments (at angle $\theta_1$)
in a periodic trajectory 
is such that $l_S\cos(\theta_1) = m_1L$, where $m_1$ is
an integer and $L$ is the size of the square. Similarly,
the total length, $l_{S'}$ of $S'$ segments in a periodic 
trajectory is such that $l_{S'}\sin(\theta_3) = n_1L$ where
$n_1$ is again an integer. It follows then that the total
length, $l_P$ of all $P$ segments in a trajectory with winding numbers
($M,N$) is such that : 

\be
l_P\cos(\theta_2) =  [(2M - m_1)L - n_1L\cot(\theta_3)]  
\ee

\noindent
On equating the total projected length of the trajectory along
the vertical edge ($l_S\sin(\theta_1) + l_P\sin(\theta_2)
+ l_{S'}\sin(\theta_3)$) to $2NL$ we obtain~: 

\be
m_1\tan(\theta_1) + [(2M - m_1) - n_1\cot(\theta_3)]\tan(\theta_2)
= (2N - n_1) \label{eq:central} 
\ee 

\noindent
where $m_1$ can vary from 0 to $2M$ while $n_1$ can vary from 0 to $2N$.
Note that $\theta_2$ and $\theta_3$ can be expressed in terms 
of $\theta_1$ so that for a given value of ($m_1,n_1$), the root
of eq.~(\ref{eq:central}) (if any) can be determined.
However not all ($m_1,n_1$) admit
real solution though in general any set of winding numbers,
$(M,N)$ allows more than one periodic orbit. The case 
$m_1 = 2M, n_1 = 0$ corresponds to a normal periodic orbit
that suffers no conversion. 

\par The total length of all $S$ ray
segments in any trajectory with winding numbers $(M,N)$ and
labelled by ($m_1,n_1$) is  $l_S + l_{S'} = 
Lm_1/\cos(\theta_1) + Ln_1/\sin(\theta_3)$
while the length of $P$ ray segments put together is
$l_P = (2M-m_1 - n_1\cot(\theta_3))L/\cos(\theta_2)$. 
The time period of the trajectory is thus $(l_S+l_{S'})/c_S + l_P/c_P$.

\par Clearly, these $S$, $S'$ and $P$ segments can be arranged in several
ways so that an orbit with a given ($M,N,m_1,n_1$) will in
general be degenerate. In order to compute the degeneracy, it
is necessary to construct symbol itineraries that are not related
by cyclic permutation. It  turns out that four distinct
symbols, {$S,S',P,P'$} together with the winding numbers
($M,N$) specify a trajectory. Here $S(P)$ and $S'(P')$  
refer to a single stretch of $S(P)$ ray 
between two consecutive vertical and horizontal edges
respectively  (see Fig. 2) though it must be noted 
that $P$ and $P'$ have the same angle $\theta_2$ where as $S$ and $S'$
are at angles $\theta_1$ and $\theta_3$ respectively. Also, the total
length of all $P$ segments is in general not equal to 
$m_2L/\cos(\theta_2) + n_2L/\sin(\theta_2)$ where $m_2$ and
$n_2$ are integers that count the number of $P$ and $P'$ segments
respectively. This is due to the fact that there exist small segments of
$P$ ray joining adjacent vertical and horizontal edges.

\par Consider for example the case with ($m_1,n_1,m_2,n_2$) = 
(1,1,1,1) \cite{note1}.
There are six distinct itineraries : 
$(S,S',P,P'), (S,S',P',P), (S,P,S',P')
(S,P,P',S'), (S,P',S',P), (S,P',P,S')$ and some 
of these correspond to periodic trajectories with
winding numbers, $(M,N) = (2,2)$ \cite{note2} .
It is implicit
however that small segments of $P$ rays joining adjacent
vertical and horizontal edges exist, whenever the following two
symbols occur consecutively in an itinerary : $(S,S'), (S',P),
(S,P')$ or $(P,P')$. Fig. 2 illustrates this for 
$(S',P,S,P')$ with $(M,N) = (2,2)$.  

\par Finally it is important to remark that the above symbols
do not specify a trajectory uniquely and it is possible that
more than one itinerary corresponds to the same orbit. This
occurs when the total length of a single $P$ segment in a 
trajectory is such that it hits at least two vertical 
and also two horizontal edges.

\par For $\kappa < \kappa_c$, the analysis gets increasingly
complicated though as $\kappa \rightarrow 1^+$, all daughter
rays born out of a parent periodic orbit at an angle
$\theta_1 = \tan^{-1}(N/M)$ are eventually periodic with
the same length (for $c_S = c_P = 1$, the time period equals the
length). The degeneracy is thus $2^{2(M + N)}$
so that the density of orbit lengths can be
expressed as 

\be
d(l) = \sum_M \sum_N \delta (l - 2L\sqrt{M^2 + N^2})\;e^{2(M+N){\rm ln}\; 2} 
\ee

\noindent
The average proliferation rate can thus be obtained by integrating
over $M$ and $N$ as :

\bea
d_{av}(l)& =& \int dM \int dN\; \delta(l - 2L\sqrt{M^2 + N^2})\;e^{2(M+N)
{\rm ln}\; 2}\\ 
         & = & \int_0^{\pi/2} 
d\theta \int dr {r\over 4L^2}\; 
\delta(l - r) e^{r \sqrt{2}\; {\rm ln}\; 2 
 \cos(\theta - \pi/4)/L} 
\eea

\noindent
where $2LM = r\cos(\theta)$ and $2LN = r\sin(\theta)$. 
The $r$ integration
is trivial and the $\theta$ integration can be evaluated
asymptotically for large $l$ using the Laplace method \cite{erdelyi}.
We finally obtain~:

\be
d_{av}(l) = {l\over 4L^2} \sqrt{\pi\over 2hl}\; e^{hl}
\ee

\noindent 
where the growth exponent $h = {\sqrt{2}\over L}{\rm ln}\;2$. 
For $L = \pi/4$, $h = 1.248$.  

\par
As $\kappa \rightarrow \infty$, no conversion is allowed 
and the average density of periodic orbits is then

\be
d_{av}(l) = {2\pi l\over 16L^2}
\ee 
  
With this background, we now present some numerical
results on the proliferation rate of periodic orbits
for $\kappa > \kappa_c$.

\section{Numerical Results}
\label{sec:numerics}

\ We first consider the case, $\kappa = 1.429$ and choose
$c_S = 1$   and $c_P = \kappa$.
Using the procedure outlined above, we have generated
all periodic orbits that have time periods less that $7.5$. 
We then construct the staircase function, $N(T) = \sum_i \Theta(T
- T_i)$ which counts the number of periodic orbits with
$T_i \leq T$. 
Fig.~4 shows a plot of ${\rm ln}\;N(T)$ as a function
of $T$ together with the best fitting straight line.
The fit
is good indicating an exponential proliferation
of periodic orbits. The slope, which is a measure
of the growth rate, equals 1.509.

\par Note that a 
direct comparison with the growth rate  at $\kappa \rightarrow 1^+$
(obtained in section~\ref{sec:periodic})
is not possible due to the fact that the velocities at
$\kappa = 1.429$ are necessarily 
different. Assuming that all conversions are allowed, the
mean velocity required to achieve a growth rate 
of $1.5$ is about $1.2$. 

\par We next consider the cases when $\kappa$ equals $8$ 
and $30$ and choose 
$c_S = 1/\kappa$ and $c_P = 1$. The
range of angles which permit conversion at $\kappa = 8$ 
is now much smaller 
and a large fraction of periodic orbits do not undergo 
any conversion for the time periods considered.
The proliferation rate is thus sub-exponential (for
a plot, see Fig.~5).

\par
For $\kappa = 30$, conversion is further
inhibited as the range of angles (in which conversion
is allowed) becomes smaller still.
This is evident in Fig.~5 where we plot ${\rm ln}\;N(T)$
as a function of ${\rm ln}\;T$ for $\kappa = 8$ and $30$.
The curve for $\kappa = 30$ has a good linear fit with
slope $2.22$ indicating 
a power law behaviour (note that the exponent is expected to 
be $2$ in the limit $\kappa \rightarrow \infty$).
The curve for $\kappa = 8$
follows this till ${\rm ln}\;T \simeq 2$ and then 
increases as more converted orbits are included.

\par This substantiates our analysis and shows
that with a decrease in the average number of
conversions as $\kappa$ increases, the proliferation
law for short periods shows a crossover from
exponential to power law behaviour. However for large
but finite $\kappa$, exponential proliferation
is eventually expected to dominate for large $T$
though these orbits are not easily accessible to
computations.

\section{Discussions and Conclusions}
\label{sec:conc}

\par We have, in the preceding sections, analysed a simple
ray-splitting system and shown that as the parameter
$\kappa$ increases from $1^+$, mode conversion is 
progressively inhibited and does not take place  
with every reflection at the boundary. This shows
up in the proliferation law of periodic orbits
as well. Surprisingly, exponential proliferation
persists for $\kappa \simeq \sqrt{2}$ where the number
of directions accessible to the daughter rays is only
three. For short periods,
there is a crossover to sub-exponential proliferation 
with increasing $\kappa$ since conversion can occur
in a progressively shrinking range of angles that
are (nearly)
parallel to the two edges and converted periodic
orbits are thus longer on an average.

\par In terms of conservation laws, the decay in
intensity that accompanies splitting must be 
compensated by a faster proliferation of
periodic orbits, though not necessarily
of the exponential kind. The sum rule for
ray-splitting however needs to be derived in
order to be more specific, though heuristically,
its form should be similar to Eq.~(\ref{eq:consv})
with each term having an additional factor 
representing the intensity loss. This is an
important area to explore for sum rules can
be put to practical use for example in 
checking whether all periodic orbits up to 
a certain length have been determined.

\par Our computations were limited by the fact
that a set of four symbols together with the winding
numbers were necessary to label periodic orbits.
We have, in each case included all periodic orbits
with symbols strings of length 10 and the period $T$
up to which all periodic orbits are available is decided
by the shortest orbit with symbol string of length greater than
$10$. For boundary shapes leading to hyperbolicity, the
complexities increase making longer  
orbits practically inaccessible to computations. Thus, 
even though our results for larger values of $\kappa$ are
limited to short orbits, they are significant in this 
light.

\section{Acknowledgements}

\par The author acknowledges valuable help from Mark Oxborrow 
and Per Rosenqvist and several useful discussions with 
Bertrand Georgeot,  
Predrag Cvitanovi\'{c}, Gregor Tanner and Niall Whelan.

\nopagebreak

\begin{figure}
\end{figure}
\vskip 2.25 in
\setlength{\unitlength}{0.240900pt}
\ifx\plotpoint\undefined\newsavebox{\plotpoint}\fi
\sbox{\plotpoint}{\rule[-0.200pt]{0.400pt}{0.400pt}}%

\vskip 0.5 in

\begin{figure}
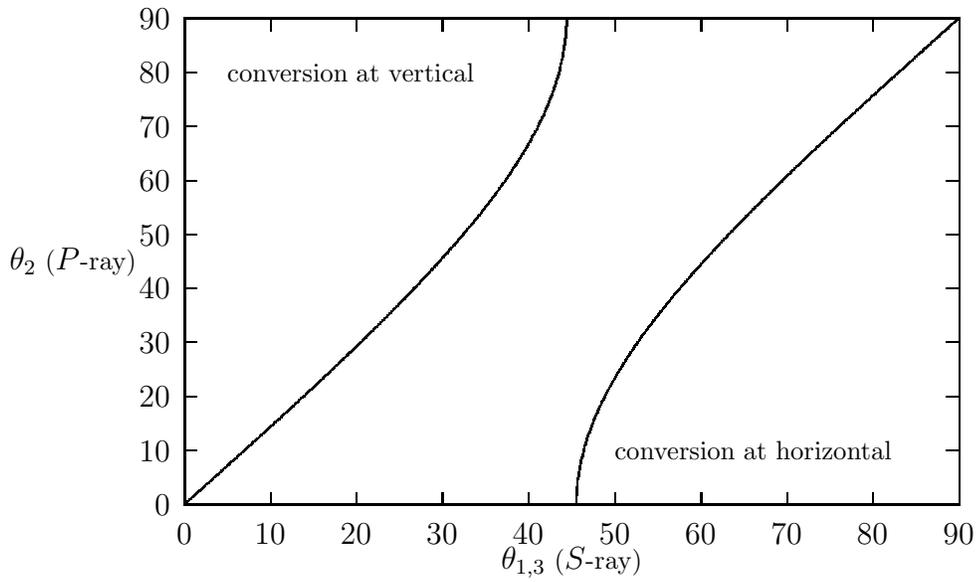


\caption{The map shows the relationship between the angles of
the S and P ray on conversion at $\kappa = 1.429$.
All angles are measured with respect to the X
axis; i.e. horizontal edge. Notational details can be found in the
text.}
\end{figure}
\pagebreak
$\;$
\vskip .2 in

\setlength{\unitlength}{0.240900pt}
\ifx\plotpoint\undefined\newsavebox{\plotpoint}\fi
%
\vskip 0.05 in
\begin{figure}
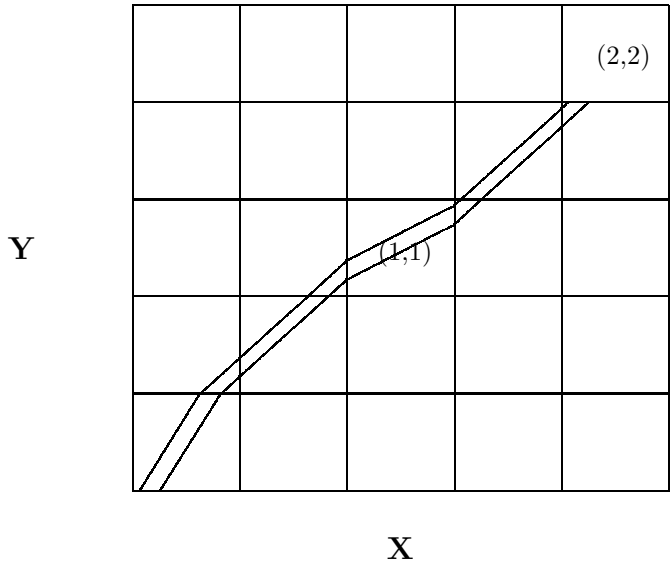

\caption{Two unfolded periodic orbits belonging to the same family
with winding number (2,2) and symbol itinerary ($S',P,S,P'$).
Note the small segments of $P$ ray between $S'$ and $P$ as well
as $S$ and $P'$. } 
\end{figure}

\vskip 0.2 in

\setlength{\unitlength}{0.240900pt}
\ifx\plotpoint\undefined\newsavebox{\plotpoint}\fi
\sbox{\plotpoint}{\rule[-0.200pt]{0.400pt}{0.400pt}}%
%
\vskip 0.05 in
\begin{figure}
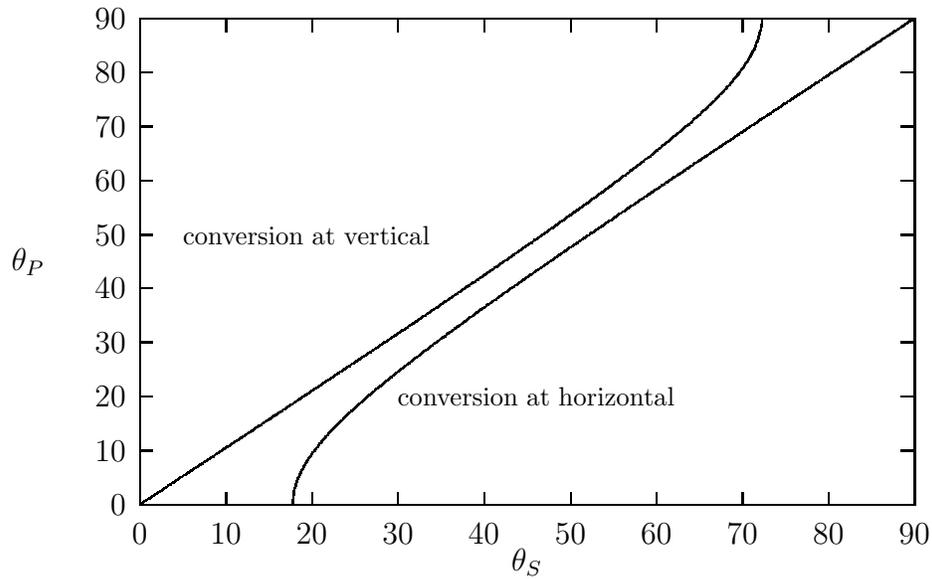

\caption{The map shows the conversions possible at $\kappa = 1.05$.
The angles $\theta_S$ and $\theta_P$ denote respectively the
angles of the $S$ and $P$ ray measured with respect to the
$X$ axis (horizontal edge). Note that there is a range of angles
where conversion can occur at both the horizontal
and vertical edges. This is referred to as an {\it overlap}.}
\end{figure}
\vskip 0.2 in

\setlength{\unitlength}{0.240900pt}
\ifx\plotpoint\undefined\newsavebox{\plotpoint}\fi
\sbox{\plotpoint}{\rule[-0.200pt]{0.400pt}{0.400pt}}%

\vskip 0.05 in
\begin{figure}
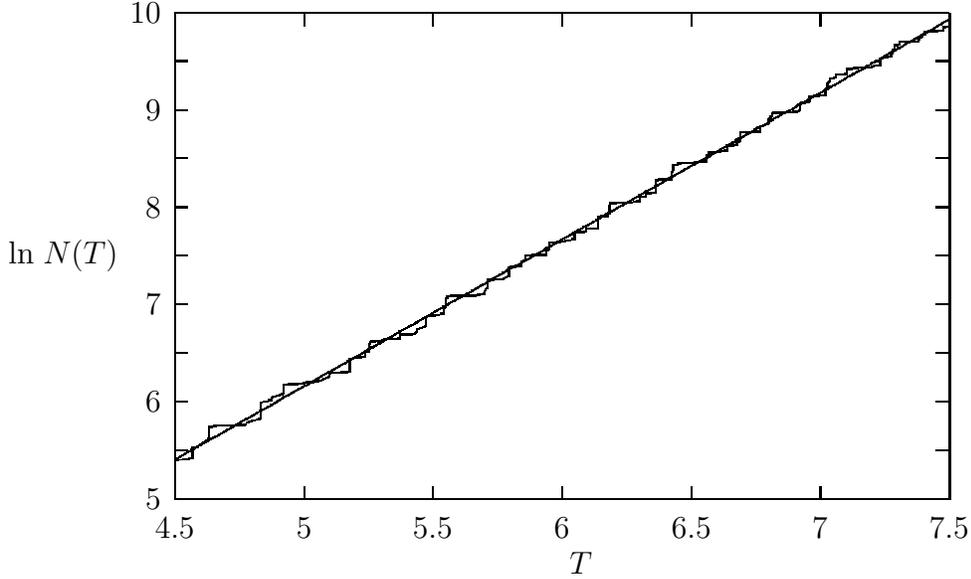

\caption{A plot of ${\rm ln}\; N(T)$ as a function of $T$ together with the
best fitting straight line. Here
$N(T)$ counts the number of periodic orbits that suffer conversion
and have periods less than or equal to $T$.
The growth exponent as derived from the slope is 1.509.}
\end{figure}

\vskip 0.2 in

\setlength{\unitlength}{0.240900pt}
\ifx\plotpoint\undefined\newsavebox{\plotpoint}\fi
\sbox{\plotpoint}{\rule[-0.200pt]{0.400pt}{0.400pt}}%
%
\vskip 0.05 in
\begin{figure}
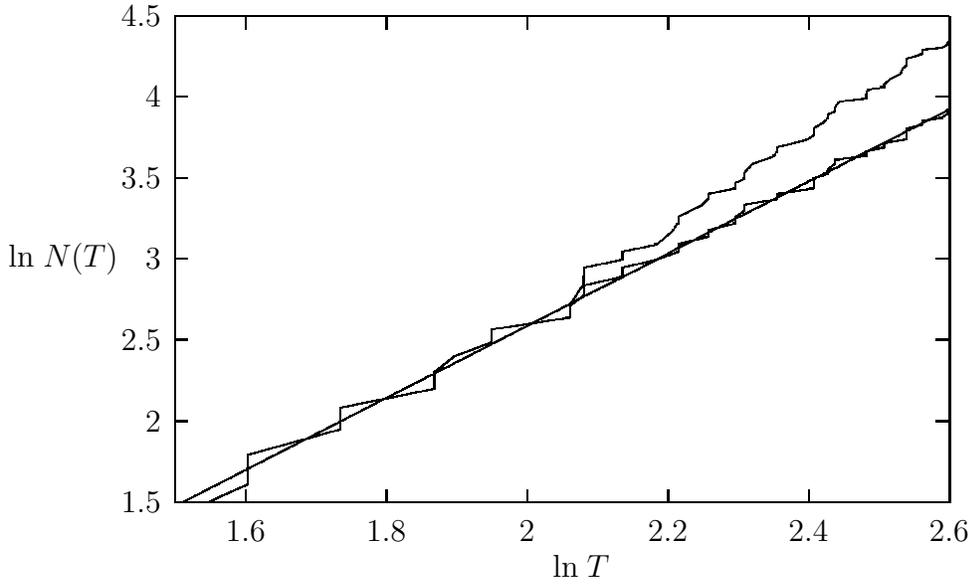

\caption{The proliferation law for $\kappa = 8$ (upper curve)
and $\kappa = 30$ (lower curve). Note that we plot ${\rm ln}\;N(T)$
as a function of ${\rm ln}\;T$. Also shown is the best fitting
straight line for $\kappa = 30$.}

\end{figure}


\begin{thebibliography}{99}
\bibitem{fnote1} Other elastic constants such as Young's
modulus, Poisson's ratio and the Bulk modulus can be expressed
in terms of the Lam\'{e} constants.
\bibitem{note1} $m_2$ and $n_2$ can vary between 0 and
[$l_P\cos(\theta_2)/L$] or [$l_P\sin(\theta_2)/L$]
respectively.
\bibitem{note2} In general two trajectories with different
(M,N) can have the same ($m_1,n_1,m_2,n_2$).
\end{thebibliography}
\end{document}